\documentclass[12pt]{iopart}

\usepackage[dvips]{graphicx}

\begin{document}

\title[Evolving networks consist of cliques]{Evolving networks consist of cliques}

\author{Zhongzhi Zhang}
\address{Department of Computer Science and Engineering, Fudan
University, Shanghai 200433, China}
\address{Shanghai Key Lab of Intelligent Information Processing,
Fudan University, Shanghai 200433, China}
\ead{zhangzz@fudan.edu.cn}

\author{Shuigeng Zhou}
\address{Department of Computer Science and Engineering, Fudan
University, Shanghai 200433, China}
\address{Shanghai Key Lab of Intelligent Information Processing,
Fudan University, Shanghai 200433, China}
\ead{sgzhou@fudan.edu.cn}

\begin{abstract}
Many real networks have cliques as their constitutional units. Here
we present a family of scale-free network model consist of cliques,
which is established by a simple recursive algorithm. We investigate
the networks both analytically and numerically. The obtained
analytical solution shows that the networks follow a power-law
degree distribution, with degree exponent continuously tuned between
2 and 3, coinciding with the empirically found results. The exact
expression of clustering coefficient is also provided for the
networks. Furthermore, the investigation of the average path length
reveals that the networks possess small-world feature.
\end{abstract}



\maketitle


\section{Introduction}
Over the last few years, it has been suggested that a lot of social,
technological, biological, and information networks share the
following three striking statistical characteristics
\cite{AlBa02,DoMe02,Ne03,DoMe03,SaVe04,NeBaWa06,BoLaMoChHw06}:
power-law degree distribution \cite{BaAl99}, high clustering
coefficient \cite{WaSt98}, and small average path length (APL).
Power-law degree distribution indicates that the majority of nodes
in such networks have only a few connections to other nodes, whereas
some nodes are connected to many other nodes in the network. Large
clustering coefficient implies that nodes having a common neighbor
are far more likely to be linked to each other than are two nodes
selected randomly. Short APL shows that the expected number of links
needed to pass from one arbitrarily selected node to another one is
low, that is, APL grows logarithmically with the number of nodes or
slower.

Mimicking such complex real-life systems is an important issue. A
wide variety of models have been
proposed~\cite{AlBa02,DoMe02,Ne03,DoMe03,SaVe04,NeBaWa06,BoLaMoChHw06},
among which the most well-known successful attempts are the Watts
and Strogatz's (WS) small-world network model \cite{WaSt98} and
Barab\'asi and Albert's (BA) scale-free network model \cite{BaAl99},
which have attracted an exceptional amount of attention from a wide
circle of researchers and started an avalanche of research on the
models of systems within the physics community. After that, a
considerable number of other models and mechanisms, which may
represent processes more realistically taking place in real-life
systems, have been developed. These include nonlinear preferential
attachment~\cite{KaReLe00}, initial attractiveness~\cite{DoMeSa00},
edge rewiring~\cite{AlBa00} and removal~\cite{DoMe00b}, aging and
cost~\cite{AmScBaSt00}, competitive dynamics~\cite{BiBa01},
duplication~\cite{ChLuDeGa03}, weight~\cite{WWX1,WWX2}, geographical
constraint~\cite{RoCoAvHa02,ZhRoGo05,ZhRoCo05a}, Apollonian
packing~\cite{AnHeAnSi05,DoMa05,ZhCoFeRo05,ZhYaWa05,ZhRoCo05,ZhRoZh06}
and so forth.

The above mentioned models and mechanisms may provide valuable
insight into some particular real-life networks. However, different
networks have different creating mechanisms, it is almost impossible
to mimic all real-life systems based on several special models.
Thus, it is necessary that we should model peculiar networks
according to their corresponding generating mechanisms.

In real-life world, many networks consist of cliques. For example,
in movie actor collaboration network \cite{WaSt98} and science
collaborating graph~\cite{Ne01a}, actors acting in the same film or
authors signing in the same paper form a clique, respectively. In
corporate director network~\cite{BaCa04}, directors as members in
the same board constitute a clique. Analogously, in public transport
networks~\cite{SiHo05}, bus (tramway, or underground) stops shape a
clique if they are consecutive stops on a route, and in the network
of concepts in written texts~\cite{CaLoAnNeMi06}, words in each
sentence in the text is added to the network as a clique. All these
pose a very interesting and important question of how to build
evolution models based on this particularity of network
component---cliques.

In this paper, we suggest a growing evolution network model with
cliques as its basic constitutional units, giving high general
versatility for growth mechanisms. The model is governed by three
tunable parameters $p$, $q$, and $m$, which control the relevant
network characteristics. Our networks have a power-law degree
distribution with degree exponent changeable between 2 and 3, a very
large clustering coefficient, and a small-world feature. The
proposed model considers systematic reorganization of cliques as its
building block, which is helpful for understanding development
processes and controls in real-world networks.

\section{Network construction}
\begin{figure}
\begin{center}
\includegraphics[width=12cm]{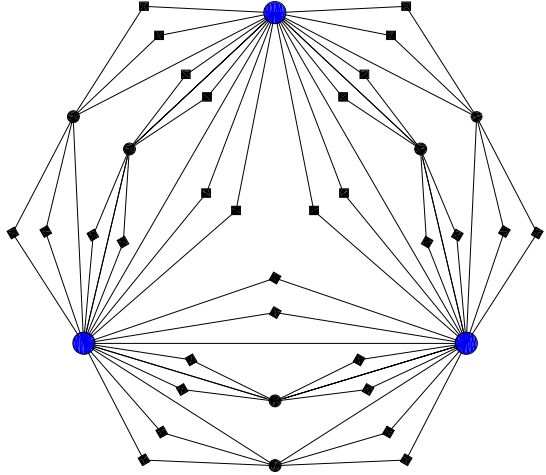}
\caption{Illustration of a deterministically growing network in the
case of $p=1$, $q=2$, and $m=2$, showing the first three steps of
growing process. } \label{network}
\end{center}
\end{figure}

We construct the networks in a recursive manner and  denote the
networks after $t$ generations by $Q(q,t)$, $q\geq 2, t\geq 0$.
Figure~\ref{network} shows the network growing process for a
particular case of $p=1$, $q=2$, and $m=2$. The networks are
constructed as follows: For $t=0$, $Q(q,0)$ is a complete graph
$K_{q+1}$ (or $(q+1)$-clique). For $t\geq 1$, $Q(q,t)$ is obtained
from $Q(q,t-1)$. For each of the existing subgraphs of $Q(q,t-1)$,
with probability $p$ $(0<p\leq 1)$, $m$ ($m$ is a positive integer)
new vertices are created, and each is connected to all the vertices
of this subgraph. The growing process is repeated until the network
reaches a desired order.

There are at least three limiting cases of our model listed below.
(i) When $q=2$, $p=1$, and $m=1$, the networks are exactly the same
as the pseudofractal scale-free web~\cite{DoGoMe02}. (ii)  When
$q=2$, $p\rightarrow0$ (but $p \neq0$), and $m=1$, our model is
reduced to the scale-free network with size-dependent degree
distribution~\cite{DoMeSa01}. (iii) When $q=2$, $0<p\leq 1$, and
$m=1$, our networks coincide with the stochastically growing
scale-free network described in Ref. \cite{Do03}. (iv) When $q\geq
2$, $p=1$, and $m=1$, our networks reduce to the recursive graphs
discussed in Ref. \cite{CoFeRa04}.

Next we compute the numbers of nodes (vertices) and links (edges) in
$Q(q,t)$. Let $L_v(t)$, $L_e(t)$ and $K_{q,t}$ be the numbers of
vertices, edges and $q$-cliques created at step $t$, respectively.
Note that the addition of each new node leads to $q$ new $q$-cliques
and $q$ new edges. So, we have $L_e(t)=K_{q,t}=qL_v(t)$. Then, at
step 1, we add expected $L_v(1)=mp(q+1)$ new nodes and
$L_e(1)=mpq(q+1)$ new edges to $Q(q,0)$. After simple calculations,
one can obtain that at $t_i$($t_i>1$) the numbers of newly born
nodes and edges are $L_v(t_i)=mp(q+1)(1+mpq)^{t_i-1}$ and
$L_e(t_i)=mpq(q+1)(1+mpq)^{t_i-1}$, respectively. Thus the average
number of total nodes $N_t$ and edges $E_t$ present at step $t$ is
\begin{eqnarray}\label{Nt1}
N_t=\sum_{t_i=0}^{t}n_v(t_i)=\frac{(q+1)[(mpq+1)^{t}+q-1]}{q}
\end{eqnarray}
and
\begin{eqnarray}\label{Et1}
E_t=\sum_{t_i=0}^{t}n_e(t_i)=\frac{(q+1)[2(mpq+1)^{t}+(q-2)]}{2},
\end{eqnarray}
respectively. So for large $t$, The average degree $\overline{k}_t=
\frac{2E_t}{N_t}$ is approximately $2q$.

\section{Topological properties}
Topology properties are of fundamental significance to understand
the complex dynamics of real-life systems. Here we focus on three
important characteristics: degree distribution, clustering
coe¡Àcient, and average path length, which are determined by the
tunable model parameters $p$, $q$, and $m$.

\subsection{Degree distribution}
\begin{figure}
\begin{center}
\includegraphics[width=9cm]{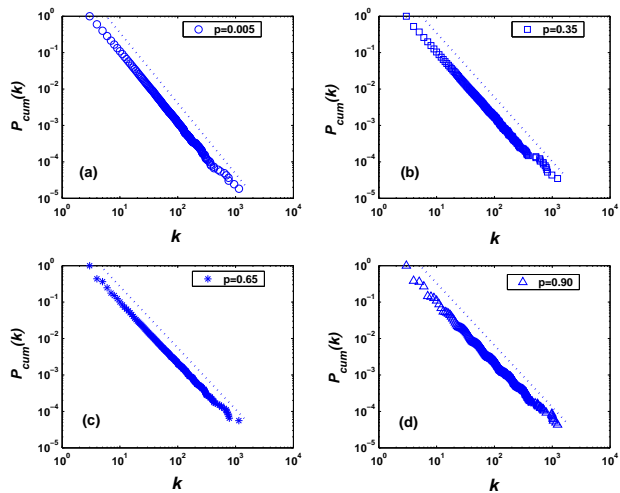}
\caption{The cumulative degree distribution $P_{cum}(k)$ at various
$p$ values for the case of $q=2$ and $m=1$. The circles (a), squares
(b), stars (c), and triangles (d) denote the simulation results for
networks with different evolutionary steps $t=1350$, $t=25$, $t=16$,
and $t=13$, respectively. The four straight lines are the
theoretical results of $\gamma(d,q)$ as provided by equation (6).
All data are from the average of 50 independent runs.} \label{Fig3}
\end{center}
\end{figure}

When a new node $i$ is added to the networks at step $t_i$, it has
degree $q$ and forms $q$ $q$-cliques. Let $L_q(i,t)$ be the number
of $q$-cliques at step $t$ that will possibly created new nodes
connected to the node $i$ at step $t+1$. At step $t_i$, $L_q(i,
t_i)=q$. By construction, we can see that in the subsequent steps
each new neighbor of $i$ generated $q-1$ new $q$-cliques with $i$ as
one vertex of them.  Then at step $t_i+1$, there are $mpq$ new nodes
which forms $mpq(q-1)$ new $q$-cliques containing $i$.  Let $k_i(t)$
be the degree of $i$ at step $t$. We can easily find following
relations for $t>t_i+1$:
\begin{equation}
\Delta k_i(t)=k_i(t)-k_i(t-1)=mpL_q(i,t-1)
\end{equation}
and
\begin{equation}
L_q(i,t)=L_q(i,t-1)+(q-1)\Delta k_i(t).
\end{equation}
From the above two equations, we can derive: $L_q(i, t+1)= L_q(i,
t)[1+mp(q-1)]$. Since $L_q(i, t_i)=q$, we have $L_q(i,
t)=q[1+mp(q-1)]^{t-t_i}$ and $\Delta
k_i(t)=mpq[1+mp(q-1)]^{t-t_i-1}$. Then the degree $ k_i(t)$ of node
$i$ at time $t$ is
\begin{eqnarray} \label{Ki1}
k_i(t)=k_i(t_i)+\sum_{t_h=t_i+1}^{t}{\Delta
k_i(t_h)}=q\left(\frac{[1+mp(q-1)]^{t-t_i}+q-2}{q-1}\right).
\end{eqnarray}
Since the degree of each node has been obtained explicitly as in
Eq.~(\ref{Ki1}), we can get the degree distribution via its
cumulative distribution \cite{Ne03}, i.e., $P_{cum}(k) \equiv
\sum_{k^\prime \geq k} N(k^\prime,t)/N_t \sim k^{1-\gamma}$, where
$N(k^\prime,t)$ denotes the number of nodes with degree $k^\prime$.
The detailed analysis is given as follows. For a degree $k$
\begin{equation*}
k=q\left(\frac{[1+mp(q-1)]^{t-s}+q-2}{q-1}\right),
\end{equation*}
there are  $L_v(s)=mp(q+1)(1+mpq)^{s-1}$ nodes with this exact
degree, all of which were born at step $s$. All nodes born at time
$s$ or earlier have this or a higher degree. So we have
\begin{eqnarray}
\sum_{k' \geq k}
N(k',t)=\sum_{a=0}^{s}L_v(a)=\frac{(q+1)[(mpq+1)^{s}+q-1]}{q}\nonumber.
\end{eqnarray}
As the total number of nodes at step $t$ is given in Eq.~(\ref{Nt1})
we have
\begin{eqnarray}
\left
[q\left(\frac{[1+mp(q-1)]^{t-s}+q-2}{q-1}\right)\right]^{1-\gamma}=\frac{\frac{(q+1)[(mpq+1)^{s}+q-1]}{q}}{\frac{(q+1)[(mpq+1)^{t}+q-1]}{q}}\nonumber.
\end{eqnarray}
Therefore, for large $t$ we obtain
\begin{equation*}
\left[[1+mq(q-1)]^{t-s}\right]^{1-\gamma}=(1+mpq)^{s-t}
\end{equation*}
and
\begin{equation}\label{gamma}
\gamma \approx 1+\frac{\ln (1+mpq)}{\ln[1+mp(q-1)]}.
\end{equation}
Thus, the degree exponent $\gamma$ is a continuous function of $p$
$q$, and $m$, and belongs to the interval [2,3]. For any fixed $q$,
as $p$ decrease from 1 to 0, $\gamma$ increases from $1+\frac{\ln
(1+mq)}{\ln [1+m(q-1)]}$ to $2+\frac{1}{m(q-1)}$ (see Appendix A for
the theoretic calculation of distribution for the particular case of
$m=1$). In the case $q=2$, $\gamma$ can be tunable between
$1+\frac{{\ln}3}{\ln2}$ and 3. In some limiting cases, Eq.
(\ref{gamma}) recovers the results previously obtained in
Refs.~\cite{DoGoMe02,DoMeSa01,Do03,CoFeRa04}. Figure~\ref{Fig3}
shows, on a logarithmic scale, the scaling behavior of the
cumulative degree distribution $P_{cum}(k)$ for different values of
$p$ in the case of $q=2$ and $m=1$. Simulation results agree very
well with the analytical ones.

\subsection{Clustering coefficient}

In the network if a given node is connected to $k$ nodes, defined as
the neighbors of the given node, then the ratio between the number
of links among its neighbors and the maximum possible value of such
links $k(k-1)/2$ is the clustering coefficient of the given
node~\cite{WaSt98}. The clustering coefficient of the whole network
is the average of this coefficient over all nodes in the network,
and can take on values between 0 and 1, the latter corresponding to
a maximally clustered network where all neighbors of a node are
linked to one another.

For our networks, the analytical expression of clustering
coefficient $C(k)$ for a single node with degree $k$ can be derived
exactly. When a node is created it is connected to all the nodes of
a $q$-clique, in which nodes are completely interconnected. So its
degree and clustering coefficient are $q$ and 1, respectively. In
the following steps, if its degree increases one by a newly created
node connecting to it, then there must be $q-1$ existing neighbors
of it attaching to the new node at the same time. Thus for a node of
degree $k$, we have
\begin{equation}\label{Ck}
C(k)= {{{q(q-1)\over 2}+ (q-1)(k-q)} \over {k(k-1)\over 2}}=
\frac{2(q-1)(k-\frac{q}{2})}{k(k-1)},
\end{equation}
which depends on both $k$ and $q$. For $k \gg q$, the $C(k)$ is
inversely proportional to degree $k$. The scaling $C(k)\sim k^{-1}$
has been found for some network
models~\cite{AnHeAnSi05,DoMa05,ZhCoFeRo05,ZhYaWa05,ZhRoCo05,ZhRoZh06,DoGoMe02,DoMeSa01,Do03,CoFeRa04,RaBa03},
and has also been observed in several real-life
networks~\cite{RaBa03}.

Using Eq. (\ref{Ck}), we can obtain the clustering $\overline{C}_t$
of the networks at step $t$:
\begin{equation}\label{ACCk}
\overline{C}_t=
    \frac{1}{N_{t}}\sum_{r=0}^{t}
    \frac{2(q-1)(D_r-\frac{q}{2})L_v(r)}{D_r(D_r-1)},
\end{equation}
where the sum runs over all the nodes
and $D_r$ 
is the degree of the nodes created at step $r$, which is given by
Eq. (\ref{Ki1}).

\begin{figure}
\begin{center}
\includegraphics[width=8cm]{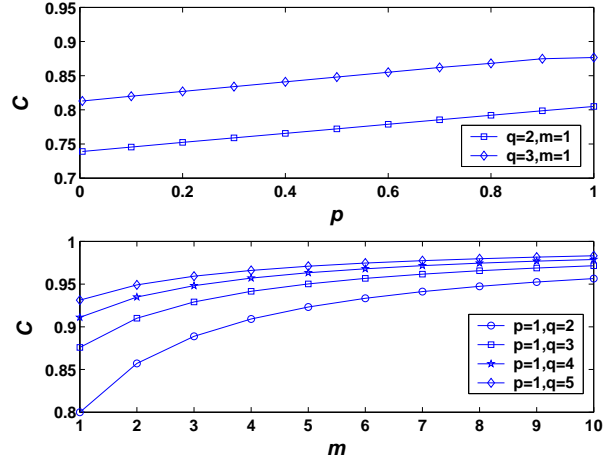}
\caption{The dependence relation of network clustering coefficient
$C$ on $p$, $q$, and $m$. Results are averaged over ten network
realizations for each datum.} \label{ACC1}
\end{center}
\end{figure}

In the infinite network order limit ($N_{t}\rightarrow \infty$), Eq.
(\ref{ACCk}) converges to a nonzero value $C$. Obviously, network
clustering coefficient $\overline{C}_t$ is a function of parameters
$p$, $q$, and $m$. If we fixed any two of them, $\overline{C}_t$
increases with the rest. Exactly analytical computation shows: in
the case $q=2$ and $m=1$, when $p$ increases from 0 to 1,
$\overline{C}$ grows from 0.739 \cite{BoPa05} to 0.8
\cite{DoGoMe02}; In the case $p=1$ and $q=2$, when $m$ increases
from 1 to infinite, $\overline{C}$ grows from 0.8 \cite{DoGoMe02} to
1; Likewise, in the case $p=1$ and $m=1$, $\overline{C}$ increases
from 0.8 to 1 when $q$ increases from 2 to infinite, with special
values $\overline{C}_t=0.8571$ and $\overline{C}_t=0.8889$ for $q=3$
and $q=4$, respectively. Therefore, the average clustering
coefficient is very large, which shows the evolving networks are
highly clustered. Figure~\ref{ACC1} exhibits the dependence of the
clustering coefficient $C$ on $p$, $q$ and $m$, which agree well
with our above conclusions.

From Figs.~\ref{Fig3} and~\ref{ACC1} and Eqs. (\ref{gamma}) and
(\ref{ACCk}), one can see that both degree exponent $\gamma$ and
clustering coefficient $\overline{C}_t$ depend on the parameter $p$,
$q$, and $m$. The mechanism resulting in this relation should be
paid further effort. The fact that a biased choice of the cliques at
each evolving step may be a possible explanation, see
Ref.~\cite{CoRobA05}.

\subsection{Average path length}
Denote the network nodes by the time step of their generations,
$v=1,2,3,\ldots,N-1,N.$ Using $L(N)$ to represent the APL of the our
model with system size $N$, then we have following realtion:
$L(N)=\frac{2\sigma(N)}{N(N-1)}$, where $\sigma(N)=\sum_{1 \leq i<j
\leq N}d_{i,j}$ is the total distance, in which $d_{i,j}$ is the
shortest distance between node $i$ and node $j$. By using the
approach similar to that in
Refs.~\cite{ZhRoCo05a,ZhYaWa05,ZhRoCo05,ZhRoZh06}, we can evaluate
the APL of the present model.

Obviously, when a new node enters the networks, the smallest
distances between existing node pairs will not change. Hence we have

\begin{equation}\label{E6}
\sigma(N+1) = \sigma(N)+ \sum_{i=1}^{N}d_{i,N+1}.
\end{equation}
Equation~(\ref{E6}) can be approximately represented as:
\begin{equation}\label{E7}
\sigma(N+1) = \sigma(N)+N+(N-q)L(N-q+1),
\end{equation}
where
\begin{equation}\label{E8}
(N-q)L(N-q+1) = {2\sigma(N-q+1) \over N-q+1} < {2\sigma(N) \over N}.
\end{equation}
Equations~(\ref{E7}) and~(\ref{E8}) provide an upper bound for the
variation of $\sigma(N)$ as
\begin{equation}\label{E9}
{d\sigma(N) \over dN} =  N + {2\sigma(N) \over N},
\end{equation}
which yields

\begin{equation}
\sigma(N) = N^2(\ln N + \omega),
\end{equation}
where $\omega$ is a constant. As $\sigma(N) \sim N^2\ln N $, we have
$L(N) \sim \ln N$.

\begin{figure}
\begin{center}
\includegraphics[width=8cm]{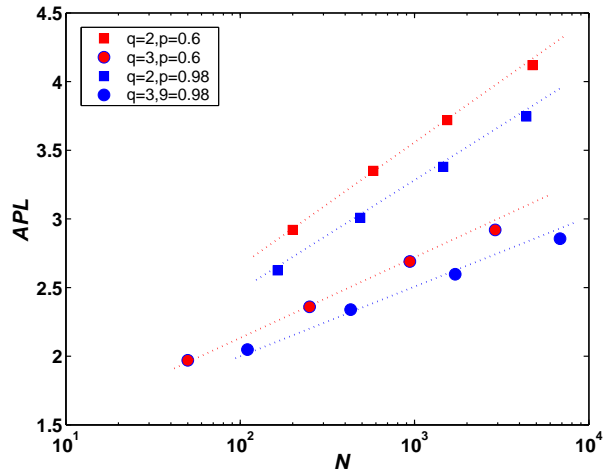} \caption{Semilogarithmic graph of the APL vs
the network size $N$ in the special case of $m=1$. Each data point
is obtained as an average of 50 independent network realizations.
The lines are linear functions of $\ln N$.} \label{Fig4}
\end{center}
\end{figure}

Note that Eq.~(\ref{E9}) was deduced from an inequality, which
implies that the increasing tendency of $L(N)$  is at most as $\ln
N$ with $N$. Thus, our model exhibits the presence of small-world
property. In Fig.~\ref{Fig4}, we show the dependence of the APL on
system size $N$ for different $p$ and $q$ in the case of $m=1$. From
Fig.~\ref{Fig4}, one can see that for fixed $q$, APL decreases with
increasing $q$; and for fixed $p$, APL is a decreasing function of
$q$. When network size $N$ is small, APL is a linear function of
$\ln N$; while $N$ becomes large, APL increases slightly slower than
$\ln N$. So the simulation results are in agreement with the
analytical prediction. It should be noted that in our model, if we
fix  $p$ and $q$, considering other values of $m$ greater than 1,
then the APL will increase more slowly than in the case $m = 1$ as
in those cases the larger $m$ is, the denser the network becomes.

Here we only give an upper bound for APL, which increases slightly
slower than $\ln N$.  Especially, in the case of $p=1$, the networks
grow deterministically, and we can compute exactly the diameter,
which is the maximum distance between all node pairs of a graph. In
this particular case, the diameter grows logarithmically with the
network size~\cite{ZhCoFeRo05,ZhRoZh06}.

\section{Conclusion}

In summary, we have proposed and studied a class of evolving
networks consist of cliques, reminiscent of modules in biological
networks or communities in social systems. We have obtained the
analytical and numerical results for degree distribution and
clustering coefficient, as well as the average path length, which
are determined by the model parameters and in accordance with large
amount of real observations. The networks are power-law, with degree
exponent adjusted continuously between 2 and 3. The clustering
coefficient of single nodes has a power-law spectra, the network
clustering coefficient is very large and independent of network
size. The intervertex separation is small, which increases at most
logarithmically as the network size. Interestingly, our networks are
formed by cliques, this particularity of the composing units may
provide a comprehensive aspect to understand some real-life systems.

\subsection*{Acknowledgment}
This research was supported by the National Natural Science
Foundation of China under Grant Nos. 60373019, 60573183, and
90612007.


\appendix

\section{Exact degree distribution for some limiting cases}
When $p\rightarrow0$ (but $p \neq0$) and $m=1$, our model turns out
to be the graph which evolves as follows (see~\cite{Do03} for
interpretation): starting with a (q+1)-clique ($t=0$), at each time
step, we choose an existing $q$-clique, then we add a new node and
join it to all the nodes of the selected $q$-clique. Note that when
$q$= 2, the particular model gives the network studied in detail in
Ref.~\cite{DoMeSa01}. Since the network size is incremented by one
with each step, here we use the step value $t$ to represent a node
created at this step. Furthermore, after a new node is added to the
network, the number of $q$-cliques increases by $q$. We can see
easily that at step $t$, the network consists of $N=t+q+1$ nodes and
$N_{q}=qN-q^{2}+1$ cliques.

One can analyze the degree distribution mathematically as follows.
Given a node, when it is born, it has degree $q$, and the number of
$q$-clique containing this node is also $q$. After that, when its
degree increases by one, the number of $q$-cliques with this node as
one of its components increases by $q-1$, so the number of
$q$-cliques for selection containing a node with degree $k$ is
$(q-1)k-q^{2}+2q$. We denote by $P_{k,N}$ the fraction of nodes with
degree $k$ when the network size is~$N$. Thus the number of such
nodes is~$NP_{k,N}$. Then the probability that the new node happens
to be connected to a particular node~$i$ having degree $k_i$ is
proportional to~$(q-1)k_i-q^{2}+2q$, and so when properly normalized
is just $[(q-1)k_i-q^{2}+2q]/(qN-q^{2}+1)$. So, between the
appearance of the $N$th and the $(N+1)$th node, the total expected
number of nodes with degree $k$ that gain a new link during this
interval is
\begin{equation}
{(q-1)k-q^{2}+2q\over qN-q^{2}+1} \times NP_{k,N} \simeq {q-1\over
q} kP_{k,N},
\end{equation}
which holds for large $N$. Observe that the number of nodes with
degree $k$ will decrease on each time step by exactly this number.
At the same time the number increases because of nodes that
previously had $k-1$ degrees and now have an extra one. Thus we can
write a master equation for the new number $(N+1)P_{k,N+1}$ of nodes
with degree $k$ thus:
\begin{equation}
(N+1)P_{k,N+1} = NP_{k,N} + {q-1\over q}
                 \left[ (k-1)P_{k-1,N} - k P_{k,N} \right].
\label{yule1}
\end{equation}
The only exception to Eq.~(\ref{yule1}) is for nodes having degree
$q$, which instead obey the equation
\begin{equation}
(N+1)P_{q,N+1} = NP_{q,N} + 1 - {q-1\over q}q P_{q,N}, \label{yule2}
\end{equation}
since by construction exactly one new such node appears on each time
step. When $N$ approaches $\infty$, we assume that the degree
distribution tends to some fixed value $P_k = \lim_{N\to\infty}
P_{N,k}$. Then from Eq.~(\ref{yule2}), we have
\begin{equation}
P_q = 1/q. \label{p1}
\end{equation}
And Eq.~(\ref{yule1}) becomes
\begin{equation}
P_k = {q-1\over q} \left[ (k-1) P_{k-1} - k P_k \right],
\end{equation}
from which we can easily obtain the recursive equation
\begin{equation}
P_k = {k-1\over k+1+\frac{1}{q-1}}\,P_{k-1},
\end{equation}
which can be iterated to get
\begin{eqnarray}
P_k &=& {(k-1)(k-2)\ldots q\over(k+1+\frac{1}{q-1})(k+\frac{1}{q-1})\ldots(q+2+\frac{1}{q-1})}\,P_q\nonumber\\
    &=& {(k-1)(k-2) \ldots (q+1)\over (k+1+\frac{1}{q-1})(k+\frac{1}{q-1})\ldots(q+2+\frac{1}{q-1})},
\end{eqnarray}
where Eq.~(\ref{p1}) has been used.  This can be simplified further
by making use of a handy property of the $\Gamma$-function,
$\Gamma(a)=(a-1)\Gamma(a-1)$ with $\Gamma(a)$ defined by:
\begin{equation}
\Gamma(a) = \int_0^\infty x^{a-1} e ^{-x} d x. \label{defsgamma}
\end{equation}  By this property and $\Gamma(1)=1$, we get
\begin{eqnarray}\label{simon}
P_k &=& \frac{(q+1+\frac{1}{q-1})(q+\frac{1}{q-1})\ldots (2+\frac{1}{q-1})}{q(q-1)\ldots 1} {\Gamma(k)\Gamma(2+\frac{1}{q-1})\over\Gamma(k+2+\frac{1}{q-1})} \nonumber\\
    &=& \frac{(q+1+\frac{1}{q-1})(q+\frac{1}{q-1})\ldots (2+\frac{1}{q-1})}{q(q-1)\ldots 1}\mathbf{B}\left(k,2+\frac{1}{q-1}\right),\nonumber\\
\label{gammas}
\end{eqnarray}
where $\mathbf{B}(a,b)$ is the Legendre beta-function, which is
defined as
\begin{equation}
\textbf{B}(a,b) = {\Gamma(a)\Gamma(b)\over\Gamma(a+b)},
\label{defsbeta}
\end{equation}
 Note that the beta-function has the interesting
property that for large values of either of its arguments it itself
follows a power law. For instance, for large $a$ and fixed~$b$,
$\textbf{B}(a,b)\sim a^{-b}$. Then we can immediately see that for
large $k$, $P_k$ also has a power-law tail with a degree exponent
\begin{equation}
\gamma = 2 + {1\over q-1}. \label{yuleexponent}
\end{equation}
For $q=2$,  $\gamma=3$, which has been obtained previously in
Ref.~\cite{Do03}.

Equation~(\ref{simon}) is similar to the Yule
distribution~\cite{Ne05} called by Simon~\cite{Simon55}. In fact,
this particular case of our model can be easily mapped into the
\emph{Yule process}, which was inspired by observations of the
statistics of biological taxa, from this perspective our model may
find applications in biological systems.


\section*{References}
\begin{thebibliography}{10}
\bibitem{AlBa02} R. Albert and A.-L. Barab\'asi,
       Rev. Mod. Phys. {\bf 74}, 47 (2002).
\bibitem{DoMe02} S.N. Dorogvtsev and J.F.F. Mendes,
Adv. Phys. {\bf 51}, 1079 (2002).

\bibitem{Ne03} M.E.J. Newman,
SIAM Review {\bf 45}, 167 (2003).

\bibitem{DoMe03}
S. N. Dorogovtsev and J. F. F. Mendes, {\it Evolution of Networks:
From Biological Nets to the Internet and WWW} (Oxford University
Press, New York, 2003).

\bibitem{SaVe04}
R. Pastor-Satorras and A. Vespignani, {\it Evolution and Structure
of the Internet: A Statistical Physics Approach} (Cambridge
University Press, Cambridge, England, 2004).

\bibitem{NeBaWa06} M. Newman, A.-L. Barab\'asi, D. J. Watts, \emph{The Structure and Dynamics of Networks}
(Princeton University Press, Princeton, 2006).

\bibitem{BoLaMoChHw06}
S. Boccaletti, V. Latora, Y. Moreno, M. Chavez and D.-U. Hwanga,
Phys. Rep. {\bf 424}, 175 (2006).


\bibitem{BaAl99} A.-L. Barab\'asi and R. Albert,
       Science {\bf 286}, 509 (1999).

\bibitem{WaSt98} D.J. Watts and H. Strogatz,
        Nature (London) {\bf 393}, 440 (1998).

\bibitem{KaReLe00}
P.L. Krapivsky, S. Redner, and F. Leyvraz,
Phys. Rev. Lett. {\bf 85}, 4629 (2000).


\bibitem{DoMeSa00}
S.N. Dorogovtsev, J.F.F. Mendes, and A.N. Samukhin,
Phys. Rev. Lett. {\bf 85}, 4633 (2000).

\bibitem{AlBa00}
R.Albert, and A.-L. Barab\'asi,
Phys. Rev. Lett. {\bf 85}, 5234 (2000).

\bibitem{DoMe00b}
S. N.Dorogovtsev, and J. F. F. Mendes,
Europhys. Lett. {\bf 52}, 33 (2000). 

\bibitem{AmScBaSt00}
L. A. N. Amaral, A. Scala, M. Barth\'el\'emy, and H. E. Stanley,
Proc. Natl. Acad. Sci. U.S.A. {\bf 97}, 11149 (2000).

\bibitem{BiBa01}
G. Bianconi, and A.-L. Barab\'asi,
Europhys. Lett. {\bf 54}, 436 (2001). 



\bibitem{ChLuDeGa03}
F. Chung, Linyuan Lu, T. G. Dewey,  D. J. Galas,
Biology {\bf 10}, 
677 (2003). 


\bibitem{WWX1} W.-X. Wang, B.-H. Wang, B. Hu, G. Yan, and Q. Ou, Phys. Rev.
Lett. \textbf{94}, 188702 (2005).

\bibitem{WWX2} W.-X. Wang, B. Hu, T. Zhou, B.-H. Wang, and Y.-B. Xie,
Phys. Rev. E \textbf{72}, 046140 (2005).

\bibitem{RoCoAvHa02}
A. F. Rozenfeld, R. Cohen, D. ben-Avraham, and S. Havlin, Phys. Rev.
Lett. \textbf{89}, 218701 (2002).

\bibitem{ZhRoGo05}
Z.Z. Zhang, L.L Rong and C.H. Guo, Physica A {\bf 363}, 567 (2006).

\bibitem{ZhRoCo05a} Z.Z. Zhang, L.L. Rong and F. Comellas,
J. Phys. A: Math. Gen. {\bf 39}, 3253 (2006).

\bibitem{AnHeAnSi05} J.S. Andrade Jr., H.J. Herrmann, R.F.S. Andrade and L.R.da Silva,
Phys. Rev. Lett. {\bf 94}, 018702 (2005).

\bibitem{DoMa05} J.P.K. Doye and C.P. Massen.
Phys. Rev. E {\bf 71}, 016128 (2005).

\bibitem{ZhCoFeRo05}
Z.Z. Zhang, F. Comellas, G. Fertin and L.L. Rong,
J. Phys. A: Math. Gen. {\bf 39}, 1811 (2006).

\bibitem{ZhYaWa05} T. Zhou, G. Yan, and B.H. Wang,
Phys. Rev. E {\bf 71}, 046141 (2005).

\bibitem{ZhRoCo05} Z.Z. Zhang, L.L Rong and F. Comellas,
Physica A {\bf 364}, 610 (2006).

\bibitem{ZhRoZh06} Z.Z. Zhang, L.L Rong and S.G. Zhou,
Phys. Rev. E {\bf 74}, 046105 (2006).

\bibitem{Ne01a}
M.E.J. Newman, Proc. Natl. Acad. Sci. U.S.A. {\bf 97}, 404 (2001).

\bibitem{BaCa04} S. Battiston and M. Catanzaro, Eur. Phys. J. B {\bf 38}, 345
(2004).

\bibitem{SiHo05}
J. Sienkiewicz and J. A. Holyst, Phys. Rev. E {\bf 72}, 046127
(2005).

\bibitem{CaLoAnNeMi06}
S.M.G. Caldeira, T.C.P. Lob{\~a}o, R.F.S. Andrade, A. Neme, and
J.G.V. Miranda1, Eur. Phys. J. B {\bf 49}, 523 (2006).

\bibitem{DoGoMe02}
S.N. Dorogovtsev, A.V. Goltsev, and J.F.F. Mendes,
          Phys. Rev. E {\bf 65}, 066122 (2002).

\bibitem{DoMeSa01}
S. N. Dorogovtsev, J. F. F. Mendes, and A. N. Samukhin, Phys. Rev. E
 {\bf 63}, 062101 (2001).

\bibitem{Do03}
S.N. Dorogovtsev, Phys. Rev. E {\bf 67}, 045102(R) (2003).


\bibitem{CoFeRa04} F. Comellas, G. Fertin and A. Raspaud,
Phys. Rev. E {\bf 69}, 037104 (2004).


\bibitem{RaBa03}
E. Ravasz and A.-L. Barab\'asi, Phys. Rev. E {\bf 67}, 026112
(2003).

\bibitem{BoPa05}
A. Barrat and R. Pastor-Satorras, Phys. Rev. E {\bf 71}, 036127
(2005).

\bibitem{CoRobA05}
F. Comellas, H. D. Rozenfeld, D. ben-Avraham, Phys. Rev. E {\bf 72},
046142 (2005).


\bibitem{Ne05} M.E.J. Newman,
Contemporary Phys.  {\bf 46}, 323 (2005).

\bibitem{Simon55}
H.~A. Simon,
Biometrika {\bf 42}, 425 (1955).


\end{thebibliography}
\end{document}